\documentclass[aps,prd,superscriptaddress,reprint,longbibliography]{revtex4-1}

\usepackage{graphicx}
\usepackage{verbatim}
\usepackage{float}
\usepackage{sidecap}
\usepackage{dcolumn}
\usepackage{bm}
\usepackage{amssymb}
\usepackage{amsmath}	
\usepackage{color}
\usepackage{multirow}
\usepackage{lipsum}
\usepackage[a4paper]{geometry}
\newgeometry{top=1.25cm,right=1.5cm,left=1.5cm,bottom=1.25cm}
\usepackage[colorlinks=true, urlcolor=magenta, pdfborder={0 0 0}]{hyperref}
\hypersetup{
linkcolor=blue,
citecolor=blue
}

\newcommand{\spss}{$sp^{3}s^{*}$ }
\newcommand{\GaBiAs}{GaBi$_{x}$As$_{1-x}$ }
\newcommand{\GaBiAsGaAs}{GaBi$_{x}$As$_{1-x}$/GaAs }

\begin{document}

\title{Atomistic tight binding study of quantum confined Stark effect in GaBi$_{x}$As$_{1-x}$/GaAs quantum wells}

\author{Muhammad Usman} \email{musman@unimelb.edu.au} \affiliation{School of Physics, The University of Melbourne, Parkville, Melbourne, 3010, Victoria Australia.} 

\begin{abstract}
\noindent
Recently, there has been tremendous research interest in novel bismide semiconductor materials (such as GaBi$_x$As$_{1-x}$) for wavelength-engineered, low-loss optoelectronic devices. We report a study of the quantum confined Stark effect (QCSE) computed for \GaBiAsGaAs quantum well (QW) structures based on large-scale atomistic tight-binding calculations. A comprehensive investigation of the QCSE as a function of the applied electric field orientations and the QW Bi fractions reveals unconventional character of the Stark shift at low Bi compositions ($x$=3.125\%). This atypical QCSE is attributed to a strong confinement of the ground-state hole wave functions due to the presence of Bi clusters. At technologically-relevant large Bi fractions ($\geq$ 10\%), the impact of Bi clustering on the electronic structure is found to be weak, leading to a quadratic Stark shift of the ground-state transition wavelength, similar to  previously observed in other conventional III-V materials. Our results provide useful insights for the understanding of the electric field dependence of the electronic and optical properties of \GaBiAsGaAs QWs, and will be important for the design of devices in the optoelectronics and spintronics areas of research.   
\end{abstract}

\maketitle

\section{Introduction}
\noindent
Quantum confined Stark effect (QCSE), which is defined as the impact of an electric field on the optoelectronic properties of nanostructures such as quantum wells (QWs) and quantum dots (QDs), has been extensively studied for a wide range of material systems~\cite{Kuo_Nature_2005, Empedocles_Science_1997, Krenner_PRL_2005, Nakaoka_PRB_2006}. In the presence of an external electrical field, the confined charge carriers (electrons and holes) feel a force which pushes the corresponding wave functions in the opposite directions. This causes a spatial separation between the charge carriers leading to the formation of an electric dipole. The QCSE plays an important role in the design of various photo-absorption and photo-emission devices such as electro-absorption modulators (EAMs), photovoltaics, and photo-diodes where the electric field mediated charge separation underpins the operation of the devices. For the conventional III-V semiconductor quantum wells such as made up of InGaAs/GaAs or InGaAs/InP materials, the QCSE leads to a quadratic red shift in the optical transition energies and wavelengths. Recently unconventional QCSE was reported for GaN/AlN quantum dots~\cite{Nakaoka_PRB_2006}, where the application of an electric field introduced a blue shift of the optical transition wavelength. The unconventional QCSE was also reported for semiconductor quantum rings exhibiting linear or quadratic shifts depending on the direction of the applied electric fields~\cite{Sousa_PRB_2017}. This work, for the first time, investigates the character of the QCSE for \GaBiAsGaAs quantum well structures and reports a peculiar character of the QCSE at low Bi fractions ($x$), where the quadratic red shift changes to linear red shift depending on the applied electric field magnitudes. Moreover, this character is found to be highly sensitive to the spatial distribution of the Bi atoms within the QW region. The detailed large-scale atomistic simulations attribute the unconventional nature of the QCSE to a strong spatial confinement of hole wave functions. By varying both $x$ and the orientation of the applied electric fields, our results indicate that the QCSE switches to a conventional quadratic red shift when the Bi fraction is increased above 10\%, which is technologically relevant for devices working at the telecomm wavelengths. The presented results offer an in-depth understanding of the QCSE for \GaBiAsGaAs quantum wells that will not only contribute to advance our knowledge on an important area of research where no prior theoretical study exists, but also provide a reliable guidance for the ongoing experimental efforts towards application of bismide alloys in the design of novel optoelectronic devices with significantly improved performance and suppressed loss mechanisms.  

Alloying the conventional III-V semiconductors such as GaAs or InAs with a dilute fraction of bismuth (Bi) ($x$) leads to a novel class of alloys known as bismides~\cite{Bismuth_containing_compounds_2013}. One of the most widely studied bismide alloy, \GaBiAs, has shown several unique properties such as a large band-gap reduction ($\approx$ 90 meV/\%Bi) with increasing $x$~\cite{Janotti_PRB_2002, Zhang_PRB_2005, Batool_JAP_2012, Usman_PRB_2011}, a strong spin orbit coupling~\cite{Broderick_SST_2012, Marko_SSMS_2014}, a cross-over between the band-gap energy (E$_g$) and the spin split-off energy ($\Delta_{SO}$) for $x \geq$ 12\%~\cite{Batool_JAP_2012, Usman_PRB_2011, Usman_PRB_2013}, and a strong confinement of the hole wave functions due to Bi pairs and clusters~\cite{Usman_APL_2014, Usman_PRB_2013, Usman_PRM_2018}. These promising characteristics of the GaBi$_x$As$_{1-x}$ material have opened new avenues for optoelectronic devices and sparked a tremendous experimental interest~\cite{Richards_JCG_2015, Balanta_JoL_2017,Mazur_JoL_2017, Aziz_APL_2017, Makhloufi_NRL_2014, Marko_SR_2016, Balanta_JPD_2016, Patil_Nanotechnology_2017}. For instance, the large band-gap engineering as a function of $x$ provides opportunities to design optoelectronic devices operating in near to mid-infrared wavelengths~\cite{Usman_PRA_2018}. The crossover between the band-gap energy and the spin split-off energy (E$_g < \Delta_{SO}$) is expected to significantly suppress CHSH Auger loss mechanisms -- the dominant source of internal losses in semiconductor laser devices~\cite{Broderick_SST_2012}. The strong spin orbit coupling has been proposed to offer benefits for spintronic applications~\cite{Mazzucato_APL_2013}. Therefore, a comprehensive understanding of the electronic structure of bismide nanostructures in response to the applied electric fields is highly desirable for the emerging optoelectronic technologies.

The application of an electric field shifts electron and hole wave functions confined in nanostructures such as QWs or QDs in opposite directions reducing their spatial overlap. The tilting of the conduction and valence band edges due to the application of potential from the electric fields leads to a decrease(increase) in the confined electron(hole) energies, inducing a red shift in the optical transition wavelength. This effect has been observed in many conventional III-V QWs and QDs, both experimentally and theoretically. Despite tremendous ongoing experimental interests and significant relevance for optoelectronic technologies, to-date, there has been no study of the QCSE for \GaBiAsGaAs QWs. In this work, we have applied a well-established and extensively bench-marked large-scale atomistic tight-binding theory to investigate the QCSE in \GaBiAsGaAs QWs as a function of both electric field orientations and Bi fractions. Our calculations revealed that at low Bi fractions ($x$=3.125\%), the QCSE exhibits unconventional character. Upon a detailed study of the confined electron and hole charge densities and their electric field dependent shifts, we attributed the atypical character of the QCSE to a very strong confinement of hole states, which require large magnitude of the electric fields to shift. Somewhat surprisingly, at large Bi fractions $x \geq$ 13\% where the ground-state transition wavelength is close to 1300 nm, the QCSE exhibits a symmetric quadratic red shift in the optical transition wavelength. This is because the presence of a large number of Bi atoms leads to a relatively uniform strain profile inside the QW region and hence the hole wave functions are more uniformly spread within the QW. Our calculations report that at $x$=13\%, the ground-state optical transition wavelength is red shifted by $\approx$22 nm for a 40 kV/cm variation in the magnitude of the applied electric field, irrespective of its orientation.

\section{Methods}
\noindent
\textbf{\textit{QW Geometry Parameters:}} A schematic diagram of the investigated quantum well structure is provided in the Fig. S1 of the supplementary material section. A single \GaBiAs quantum well with 12 nm width along the (001) direction and 20 nm lateral dimensions along the in-plane ((100) and (010)) directions is placed in GaAs material. The quantum well region consists of 196608 atoms, whereas the total size of the surrounding GaAs box is 20 nm $\times$ 20 nm $\times$ 52 nm, consisting of 851968 atoms. The boundary conditions for the simulation domain are periodic in all three spatial directions. The size of the GaAs box was selected to be sufficiently large to enable proper strain relaxation of atoms and to ensure negligible inter-well couplings along the (001) growth direction. The Bi fractions ($x$) of the QW structure are selected as 3.125\%, 6.5\%, 10\%, and 13\%, which correspond to 3072, 6389, 9831, and 12778 number of Bi atoms, respectively, randomly replacing the arsenic (As) atoms in the QW region. To investigate the effect of the lateral size of QW region, we increase the number of atoms in the QW region to 1.33 million and find that the lateral size effect on the electron and hole confinement wave functions is negligibly small. We also investigate five different random configurations to study the effect of the placement of Bi atoms as will be discussed in a later section. The ground state transition energy (GSTE) is computed from the difference between the lowest electron (e1) and the highest hole (h1) energies, and the ground state transition wavelength (GSTW) in the units of nm is defined as: GSTW = 1240/GSTE. 

\noindent
\textbf{\textit{Strain and Electronic Structure:}} The lattice mismatch between the GaAs and \GaBiAs materials leads to strain in the crystal. We compute the strain from an atomistic valence force field (VFF) method~\cite{Keating_PR_1966} by properly relaxing the crystal structure. The electronic structure calculations are performed by solving a nearest-neighbor \textit{sp$^3$s$^*$} tight-binding Hamiltonian, including spin-orbit coupling. The parameters used to model GaAs and GaBi band structures were published in our previous study~\cite{Usman_PRB_2011}. We note that our \textit{sp$^3$s$^*$} tight-binding parameters reproduce the GaBi band structure very accurately around the $\Gamma$ point, however the high energy conduction bands do not quantitatively match the published band structure, in particular along the $\Gamma$ to X direction of the Brillouin zone. This is related to the limitation of the \textit{sp$^3$s$^*$} parametrisation as discussed in the previous studies~\cite{Klimeck_SM_2000, Sawamura_OME_2018}. However, for the GaBiAs material which is a direct band-gap material, the fitted \textit{sp$^3$s$^*$} tight-binding parameters provide a very good description of the band structure properties. The model has been extensively benchmarked against a number of experimental studies on both bulk \GaBiAs materials~\cite{Usman_PRB_2011, Usman_PRB_2013} and \GaBiAsGaAs QWs~\cite{Usman_PRM_2018, Broderick_PRB_2014, Broderick_SST_2015}. Subsequently, the tight-binding model has also shown good agreement with the experimental data sets~\cite{Donmez_SST_2015, Balanta_JoL_2017, Zhang_JAP_2018, Dybala_APL_2017, Collar_AIPA_2017, Broderick_PRB_2014} as well as with the DFT calculations reported in the literature~\cite{Kudrawiec_JAP_2014, Polak_SST_2015, Bannow_PRB_2016}.

\noindent
The ground state electron and hole energies and wave functions at the $\Gamma$ point are computed for each Bi fraction in the quantum well region. The details of the VFF relaxation and tight-binding model parameters are published in our earlier work~\cite{Usman_PRB_2011}. The charge density plots shown in figures 3 and 4 are plotted as follows:

 \noindent
\begin{eqnarray}
\vert \psi_{e}(z) \vert^2 &=& \frac{1}{total \, \# \, of \, atoms \, at \, z}\sum_{all \, atoms \, at \, z} \vert \langle \psi_{e} \vert \psi_{e}  \rangle \vert^2 \\
\vert \psi_{h}(z) \vert^2 &=& \frac{1}{total \, \# \, of \, atoms \, at \, z}\sum_{all \, atoms \, at \, z} \vert \langle \psi_{h} \vert \psi_{h}  \rangle \vert^2
\label{eq:e_h_densities}
\end{eqnarray}

\noindent
where $\psi_{e}$ and $\psi_{h}$ are the ground-state electron and hole wave functions corresponding to e1 and h1 energies respectively. The $z-axis$ denotes the (001) plane. Note that in this work, charge density on anions is shown, however the charge density on cations exhibits similar trends and omitted for simplicity~\cite{Usman_APL_2014}. The computation of optical transition strengths includes charge densities on both cations and anions. The simulations are performed using NanoElectronic MOdeling (NEMO-3D) tool~\cite{Klimeck_IEEETED_2007_2}. 

\noindent
\textbf{\textit{Impact of Electric Field:}} To study the QCSE, we investigate the effect of a static electric field whose magnitude is varied between -40 kV/cm and 40 kV/cm and the orientation of electric field is selected along the three primary crystal directions: (100), (010), and (001). The impact of electric field on electronic structure is computed by including a net potential in the diagonal part of the tight-binding Hamiltonian~\cite{Usman_IOPNanotech_2011}. We note here that the  theoretical investigation of the QCSE for semiconductor nanostructures has a long history, with initial studies based on variational methods~\cite{Miller_PRL_1984, Miller_PRB_1985} and perturbation approach~\cite{Lengyel_IEEE_1990}. Recently, detailed atomistic techniques have been applied towards the computation of the QCSE~\cite{Usman_IOPNanotech_2011, Sukkabot_2018}. In this work, we have employed an atomistic tight-binding theory which is particularly well-suited to investigate the electronic structure of the highly-mismatched alloys such as \GaBiAs~\cite{Usman_PRB_2011}.    

\section{Results and Discussions}
\begin{figure}
\includegraphics[scale=0.2]{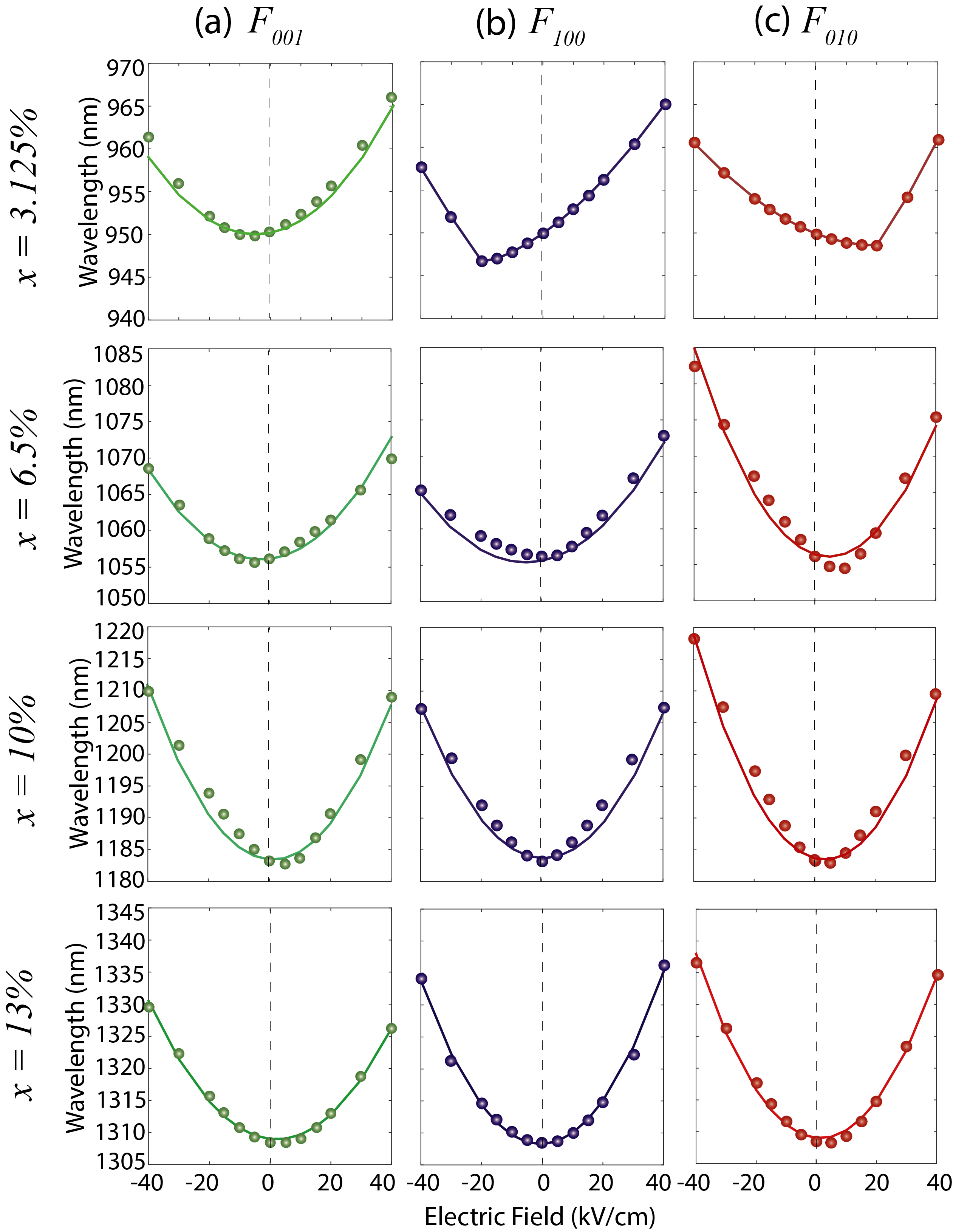}
\caption{The plots of the ground-state inter-band optical transition wavelengths are shown as a function of the Bi fractions ($x$) and the applied static electric fields ($\vec{F}_{100}$, $\vec{F}_{010}$, and $\vec{F}_{001}$) varying from -40 kV/cm to 40 kV/cm. The data points show the actual computed values and the plotted lines are the fittings from the equation 1 based on the values reported in table 1.}
\label{fig:Fig1}
\end{figure}

\noindent
\textbf{\textit{Electric Field Dependence of GSTW:}} Figure~\ref{fig:Fig1} shows the GSTW plots as a function of the applied electric fields for various Bi fractions (rows) and the three orientations of the applied electric field (columns). The application of electric field along the growth or confinement direction ($\vec{F}_{001}$) leads to a conventional Stark effect irrespective of the QW Bi fraction. The GSTW red shifts when the magnitude of the electric field increases. For a net variation of 40 kV/cm in $| \vec{F}_{001} |$, we computed a change of approximately 22 nm, 12 nm, 25 nm, and 22 nm in the GSTW for $x$ = 3.125\%, 6.5\%, 10\%, and 13\%, respectively. It is noted that the minimum value of the GSTW is not at zero electric field, which indicates that the charge carriers are not symmetrically confined in the QW region at zero electric field. It will be shown later that this asymmetry is primarily related to the asymmetric confinement of the hole wave functions, as the electron wave functions are nearly symmetrically confined along the (001) direction.

For the in-plane electric field orientations ($\vec{F}_{100}$ and $\vec{F}_{010}$), the plots of the GSTW at $x$=3.125\% showed a remarkable asymmetry with respect to the magnitude of the applied electric fields. The computed electric field dependent change in the GSTW is quadratic for -20 kV/cm $\leq \vec{F}_{100} \leq$ 40 kV/cm and -40 kV/cm $\leq \vec{F}_{010} \leq$ 20 kV/cm, and it is linear for -40 $\leq \vec{F}_{100} \leq$ -20 kV/cm and 20 $\leq \vec{F}_{010} \leq$ 40 kV/cm. In the next few sections, we will provide a detailed analysis of this atypical GSTW character by carefully analysing the electric field dependent electron and hole energies and charge density plots. Here we note that when the Bi fraction in the QW region is increased, the QCSE becomes increasingly quadratic as have been widely reported for the conventional III-V alloys. At $x$=13\%, the dependence of the QCSE on the applied electric field is completely quadratic and exhibits approximately 21 nm and 24 nm variations in the GSTW for the applied range of the $\vec{F}_{100}$ and $\vec{F}_{010}$ electric fields, respectively. These variations are similar to the $\approx$ 22 nm change in wave length for $\vec{F}_{001}$, indicating that at $x$=13\%, the QCSE is symmetric with respect to the orientation of the applied electric field. 

\begin{table*}
\caption{\label{tab:table1} The electric field dependence of the ground-state optical transition energies, GSTE, (E$_g$ = e$_1$-h$_1$) is fitted to equation 1. The fitted values of E$_g(0)$, $\alpha$ and $\beta$ are provided below for the investigated Bi fractions $x$ and the applied electric field orientations. For $x$=3.125\%, the values marked with $\dagger$ symbols are extracted for the electric field values (in the units of kV/cm) in the following range: -20 $\leq \vec{F}_{100} \leq$ 40 and -40 $\leq \vec{F}_{010} \leq$ 20, whereas the values marked with $\ddagger$ symbols are extracted for the electric field values in the following range: -40 $\leq \vec{F}_{100} \leq$ -20 and 20 $\leq \vec{F}_{010} \leq$ 40.}
\begin{tabular}{c|c|c|c|c|c|c|c|c|c|c|c|c|}
\multicolumn{13}{c}{} \\[3pt]
\cline{1-13}
\multicolumn{1}{|c|}{} & 
\multicolumn{4}{c|}{} & 
\multicolumn{4}{c|}{} & 
\multicolumn{4}{c|}{} \\[1pt]

\multicolumn{1}{|c|}{$x$} & 
\multicolumn{4}{c|}{Electric Field Orientation $\vec{F}_{001}$} & 
\multicolumn{4}{c|}{Electric Field Orientation $\vec{F}_{100}$} & 
\multicolumn{4}{c|}{Electric Field Orientation $\vec{F}_{010}$} \\[8pt] \cline{2-13}

\multicolumn{1}{|c|}{} &
\multicolumn{1}{c|}{} &
\multicolumn{1}{c|}{} &
\multicolumn{1}{c|}{} &
\multicolumn{1}{c|}{} &
\multicolumn{1}{c|}{} &
\multicolumn{1}{c|}{} &
\multicolumn{1}{c|}{} &
\multicolumn{1}{c|}{} &
\multicolumn{1}{c|}{} &
\multicolumn{1}{c|}{} &
\multicolumn{1}{c|}{} &
\multicolumn{1}{c|}{} \\[1pt] 

\multicolumn{1}{|c|}{(\%)} &
\multicolumn{1}{c|}{E$_g(0)$} &
\multicolumn{1}{c|}{$\alpha$ ($\times $10$^{-6}$)} &
\multicolumn{1}{c|}{$\beta$ ($\times $10$^{-6}$)} &
\multicolumn{1}{c|}{R$^2$} &
\multicolumn{1}{c|}{E$_g(0)$} &
\multicolumn{1}{c|}{$\alpha$ ($\times $10$^{-6}$)} &
\multicolumn{1}{c|}{$\beta$ ($\times $10$^{-6}$)} &
\multicolumn{1}{c|}{R$^2$} &
\multicolumn{1}{c|}{E$_g(0)$} &
\multicolumn{1}{c|}{$\alpha$ ($\times $10$^{-6}$)} &
\multicolumn{1}{c|}{$\beta$ ($\times $10$^{-6}$)} &
\multicolumn{1}{c|}{R$^2$} \\[8pt] \cline{1-13}

\multicolumn{1}{|c|}{} &
\multicolumn{1}{c|}{} &
\multicolumn{1}{c|}{} &
\multicolumn{1}{c|}{} &
\multicolumn{1}{c|}{} &
\multicolumn{1}{c|}{} &
\multicolumn{1}{c|}{} &
\multicolumn{1}{c|}{} &
\multicolumn{1}{c|}{} &
\multicolumn{1}{c|}{} &
\multicolumn{1}{c|}{} &
\multicolumn{1}{c|}{} &
\multicolumn{1}{c|}{} \\[1pt] 
\multicolumn{1}{|c|}{3.125} &
\multicolumn{1}{c|}{1.3127} &
\multicolumn{1}{c|}{-91.6} &
\multicolumn{1}{c|}{-17.1} &
\multicolumn{1}{c|}{0.985} &
\multicolumn{1}{c|}{1.3053$^\dagger$} &
\multicolumn{1}{c|}{-300$^\dagger$} &
\multicolumn{1}{c|}{-4$^\dagger$} &
\multicolumn{1}{c|}{0.999$^\dagger$} &
\multicolumn{1}{c|}{1.3053$^\dagger$} &
\multicolumn{1}{c|}{200$^\dagger$} &
\multicolumn{1}{c|}{-4$^\dagger$} &
\multicolumn{1}{c|}{0.999$^\dagger$} \\[8pt]
\multicolumn{1}{|c|}{} &
\multicolumn{1}{c|}{--} &
\multicolumn{1}{c|}{--} &
\multicolumn{1}{c|}{--} &
\multicolumn{1}{c|}{--} &
\multicolumn{1}{c|}{1.3258$^\ddagger$} &
\multicolumn{1}{c|}{800$^\ddagger$} &
\multicolumn{1}{c|}{0$^\ddagger$} &
\multicolumn{1}{c|}{1.0$^\ddagger$} &
\multicolumn{1}{c|}{1.3243$^\ddagger$} &
\multicolumn{1}{c|}{-800$^\ddagger$} &
\multicolumn{1}{c|}{0$^\ddagger$} &
\multicolumn{1}{c|}{1.0$^\ddagger$}  \\[8pt] \cline{1-13}

\multicolumn{1}{|c|}{} &
\multicolumn{1}{c|}{} &
\multicolumn{1}{c|}{} &
\multicolumn{1}{c|}{} &
\multicolumn{1}{c|}{} &
\multicolumn{1}{c|}{} &
\multicolumn{1}{c|}{} &
\multicolumn{1}{c|}{} &
\multicolumn{1}{c|}{} &
\multicolumn{1}{c|}{} &
\multicolumn{1}{c|}{} &
\multicolumn{1}{c|}{} &
\multicolumn{1}{c|}{} \\[1pt] 
\multicolumn{1}{|c|}{6.5} &
\multicolumn{1}{c|}{1.1735} &
\multicolumn{1}{c|}{-39.8} &
\multicolumn{1}{c|}{-8.9} &
\multicolumn{1}{c|}{0.986} &
\multicolumn{1}{c|}{1.1734} &
\multicolumn{1}{c|}{-90.7} &
\multicolumn{1}{c|}{-8.79} &
\multicolumn{1}{c|}{0.988} &
\multicolumn{1}{c|}{1.1735} &
\multicolumn{1}{c|}{143.2} &
\multicolumn{1}{c|}{-15.6} &
\multicolumn{1}{c|}{0.982} \\[8pt] \cline{1-13}

\multicolumn{1}{|c|}{} &
\multicolumn{1}{c|}{} &
\multicolumn{1}{c|}{} &
\multicolumn{1}{c|}{} &
\multicolumn{1}{c|}{} &
\multicolumn{1}{c|}{} &
\multicolumn{1}{c|}{} &
\multicolumn{1}{c|}{} &
\multicolumn{1}{c|}{} &
\multicolumn{1}{c|}{} &
\multicolumn{1}{c|}{} &
\multicolumn{1}{c|}{} &
\multicolumn{1}{c|}{} \\[1pt] 
\multicolumn{1}{|c|}{10} &
\multicolumn{1}{c|}{1.0468} &
\multicolumn{1}{c|}{35.2} &
\multicolumn{1}{c|}{-14} &
\multicolumn{1}{c|}{0.992} &
\multicolumn{1}{c|}{1.04647} &
\multicolumn{1}{c|}{-43.8} &
\multicolumn{1}{c|}{-12.7} &
\multicolumn{1}{c|}{0.991} &
\multicolumn{1}{c|}{1.0462} &
\multicolumn{1}{c|}{110.06} &
\multicolumn{1}{c|}{-16.2} &
\multicolumn{1}{c|}{0.989} \\[8pt] \cline{1-13}

\multicolumn{1}{|c|}{} &
\multicolumn{1}{c|}{} &
\multicolumn{1}{c|}{} &
\multicolumn{1}{c|}{} &
\multicolumn{1}{c|}{} &
\multicolumn{1}{c|}{} &
\multicolumn{1}{c|}{} &
\multicolumn{1}{c|}{} &
\multicolumn{1}{c|}{} &
\multicolumn{1}{c|}{} &
\multicolumn{1}{c|}{} &
\multicolumn{1}{c|}{} &
\multicolumn{1}{c|}{} \\[1pt] 
\multicolumn{1}{|c|}{13} &
\multicolumn{1}{c|}{0.9473} &
\multicolumn{1}{c|}{41.9} &
\multicolumn{1}{c|}{-8.5} &
\multicolumn{1}{c|}{0.998} &
\multicolumn{1}{c|}{0.9476} &
\multicolumn{1}{c|}{-11.7} &
\multicolumn{1}{c|}{-11.7} &
\multicolumn{1}{c|}{0.998} &
\multicolumn{1}{c|}{0.9475} &
\multicolumn{1}{c|}{31.7} &
\multicolumn{1}{c|}{-12.03} &
\multicolumn{1}{c|}{0.996} \\[8pt] \cline{1-13}

\end{tabular} 
\end{table*}

\noindent
\textbf{\textit{Built-in Dipole Moment and Polarizability:}} To provide quantitative estimates of the changes in the band gap energies as a function of the applied electric fields, we fitted the GSTE, corresponding to the GSTW plots in Fig.~\ref{fig:Fig1}, to a quadratic equation of the form: 

\begin{equation}
	\label{eq:exchange}
	E_g(\vec{F_m}) = E_g(0) + \alpha \vec{F_m} + \beta \vec{F_m}^2
\end{equation}  

\noindent
where E$_g(0)$ is the value of the GSTE in the units of eV computed without the application of electric field ($|\vec{F_m}|$=0), $m$ is the orientation of the applied electric field, $\alpha$ is the permanent dipole moment in the units of eV.cm/kV, and $\beta$ is a measure of the polarizability of the electron and hole wave functions in the units of eV.cm$^2$/kV$^2$. This equation has been widely used in the literature to investigate the electric-field-dependent shift in the GSTE~\cite{Robinson_APL_2005, Seufert_APL_2001}. The character of the GSTE shift is dependent on many factors, including the size of the nanostructures and the material properties. For example, it has been shown that the InGaN quantum dots exhibit a Stark effect which includes both linear and quadratic components~\cite{Robinson_APL_2005}; however, the Stark shift for the CdSe quantum dots shows only a quadratic dependence on the electric field~\cite{Seufert_APL_2001}. Here, we fit the equation~\ref{eq:exchange} to the computed GSTE data and show that in the case of the \GaBiAs quantum wells, the Stark effect includes both linear and quadratic components. 

Table 1 provides the fitted values of E$_g(0)$, $\alpha$ and $\beta$ for the investigated Bi fractions $x$ and the applied electric field orientations. In each case, we have also provided $R^2$ values, which indicate the quality of the fit of equation 1 to the GSTE plots. An $R^2$ value of 1.0 implies a perfect fit, indicating that the GSTE dependence on the electric field is completely represented by the quadratic formula of the equation 1. The values in the Table 1 show a general trend that for the studied \GaBiAsGaAs QWs, the value of $R^2$ increases as $x$ increases. This suggests that at larger Bi fractions, an increasingly conventional behaviour of the QCSE is computed. For $x$=3.125\%, there was not a single fit to the computed GSTE as the dependence on the in-plane electric fields ($\vec{F}_{100}$ and $\vec{F}_{010}$) varies between the linear and quadratic functions. Therefore, two separate fits were performed as indicated in the table by symbols $\dagger$ and $\ddagger$. 

From the table 1, we find that the values of $\alpha$ are larger at the small Bi fraction ($x$=3.125\%) and they generally decrease as $x$ increases. This indicates that the electron and hole wave functions are relatively largely separated at $x$=3.125\%, which gives rise to a higher built-in permanent dipole moment. This will become clearer in the later sections when we will discuss the charge density plots where the hole charge density plots show a highly confined character. 

The values of $\beta$, which is the measure of polarizability or the electric field response, increase when the Bi fraction in the QW region increases. This shows that at higher Bi fractions, the applied electric field has stronger impact on pulling the electron and hole charge carriers in the opposite directions, compared to the low Bi fractions. Therefore, a relatively large reduction in the spatial overlap between the electron and hole wave functions is expected for the QWs with Bi fractions above 10\%, which are technologically relevant~\cite{Sweeney_JAP_2013}.     

\begin{figure}
\includegraphics[scale=0.2]{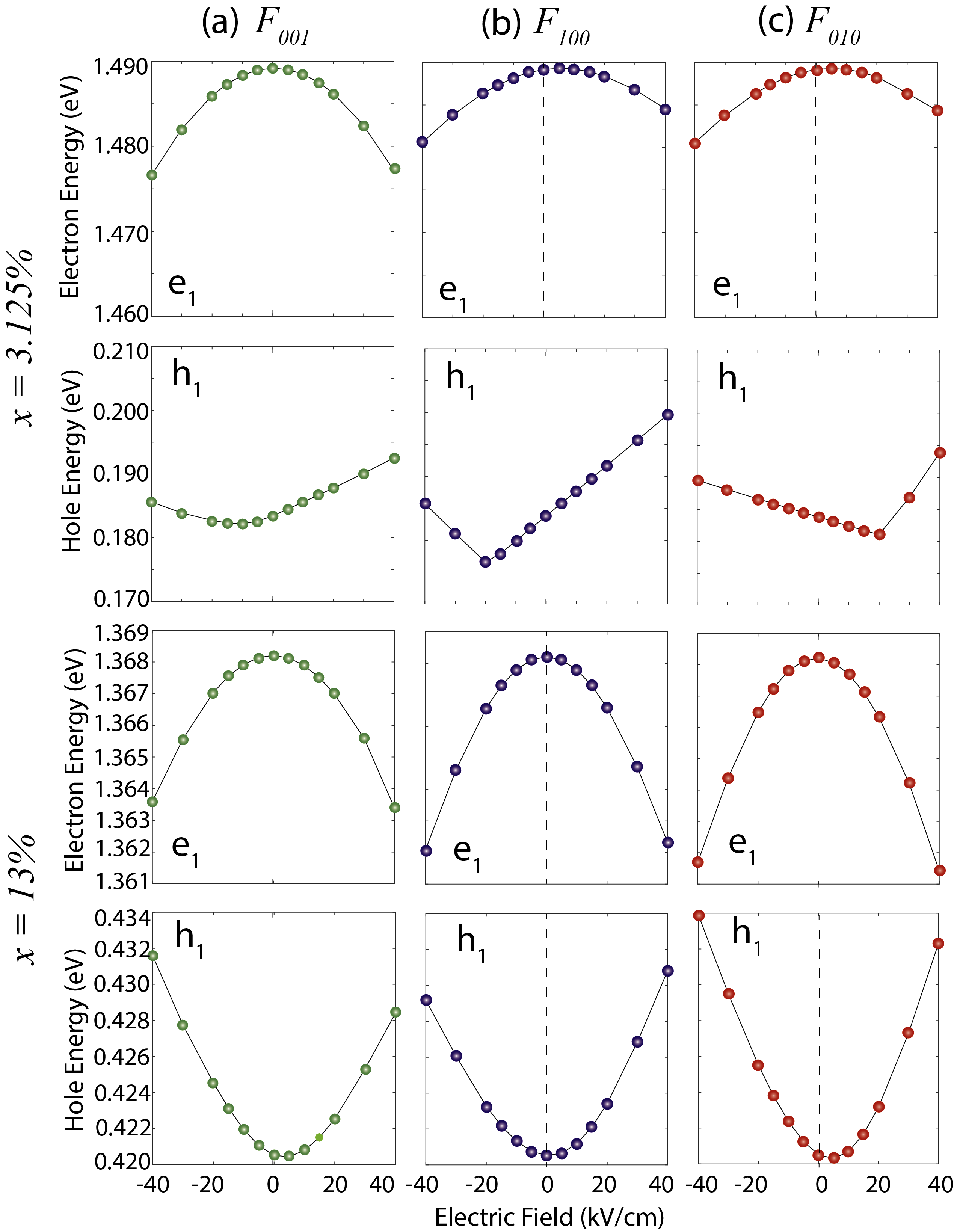}
\caption{The plot of the lowest electron (e$_1$) and the highest hole (h$_1$) energies for a \GaBiAsGaAs QW with $x$=3.125\% for different orientations of the applied electric field: (a) $\vec{F}_{001}$, (b) $\vec{F}_{100}$, and (c) $\vec{F}_{010}$.}
\label{fig:Fig2}
\end{figure}

\noindent
\textbf{\textit{Unconventional Stark Shift of Holes Energies:}} To further investigate the previously discussed unconventional character of the QCSE computed for $x$=3.125\%, we plot the lowest electron (e1) and the highest hole (h1) energies as a function of the applied electric fields in Fig.~\ref{fig:Fig2}. For comparison, we have also plotted the same energies for $x$=13\%. The corresponding plots for $x$=6.5\% and 10\% are provided in Fig. S2 of the supplementary material section for reference. From these plots, it is clearly evident that the lowest electron energies (e1) exhibit a conventional quadratic dependence on the magnitude of electric fields, irrespective of the filed orientations and Bi fractions $x$. 

The Stark shift in the highest hole energies (h1) is quite different from the Stark shift in e1 and clearly depends on the value of $x$ as well as on the orientation of the applied electric field. At large Bi fraction $x$=13\%, the hole energy Stark shift is quadratic, similar to the Stark shift in e1 energy, which explains the conventional nature of the QCSE at $x$=13\% (Fig.~\ref{fig:Fig1}). However, at $x$=3.125\%, the Stark shift in the hole energies exhibits two atypical characteristics: firstly, the computed Stark effect of the hole energies exhibits linear dependence on the electric field magnitudes. It is also noted that the minimum value of h1 is not at $|\vec{F_m}|$=0, rather it is at $\vec{F}_{001}$=-10 kV/cm, $\vec{F}_{100}$=-20 kV/cm and $\vec{F}_{010}$=20 kV/cm. Secondly, the field dependence of h1 energy is highly asymmetric on the direction of field: for -20 kV/cm $\leq \vec{F}_{100} \leq$ 40 kV/cm, the slope of h1 dependence is 0.4 meV.cm/kV and it is -0.5 meV.cm/kV for -40 kV/cm $\leq \vec{F}_{100} \leq$ -20 kV/cm. On the other hand, for -40 kV/cm $\leq \vec{F}_{010} \leq$ 20 kV/cm, the slope of h1 dependence is -0.2 meV.cm/kV and it is 0.6 meV.cm/kV for 20 kV/cm $\leq \vec{F}_{010} \leq$ 40 kV/cm. For the field direction along the growth axis, the slope is 0.2 meV.cm/kV for -5 kV/cm $\leq \vec{F}_{001} \leq$ 40 kV/cm and -0.1 meV.cm/kV for -40 kV/cm $\leq \vec{F}_{001} \leq$ -10 kV/cm. For the in-plane electric field orientations, these peculiar variations in the hole energies with larger slopes combined with relatively smaller change in the e1 energies result in the unconventional nature of the QCSE reported earlier in Fig.~\ref{fig:Fig1}. The difference in the Stark shifts of the electron and hole energies can be explained by looking at the corresponding charge densities. In the next two sections, we will first plot the spatial distributions of the electron and hole charge densities and highlight the nature of their confinement inside the QW region. We will then discuss the dependence of the charge densities on the application of the electric fields.  

\begin{figure*}
\includegraphics[scale=0.28]{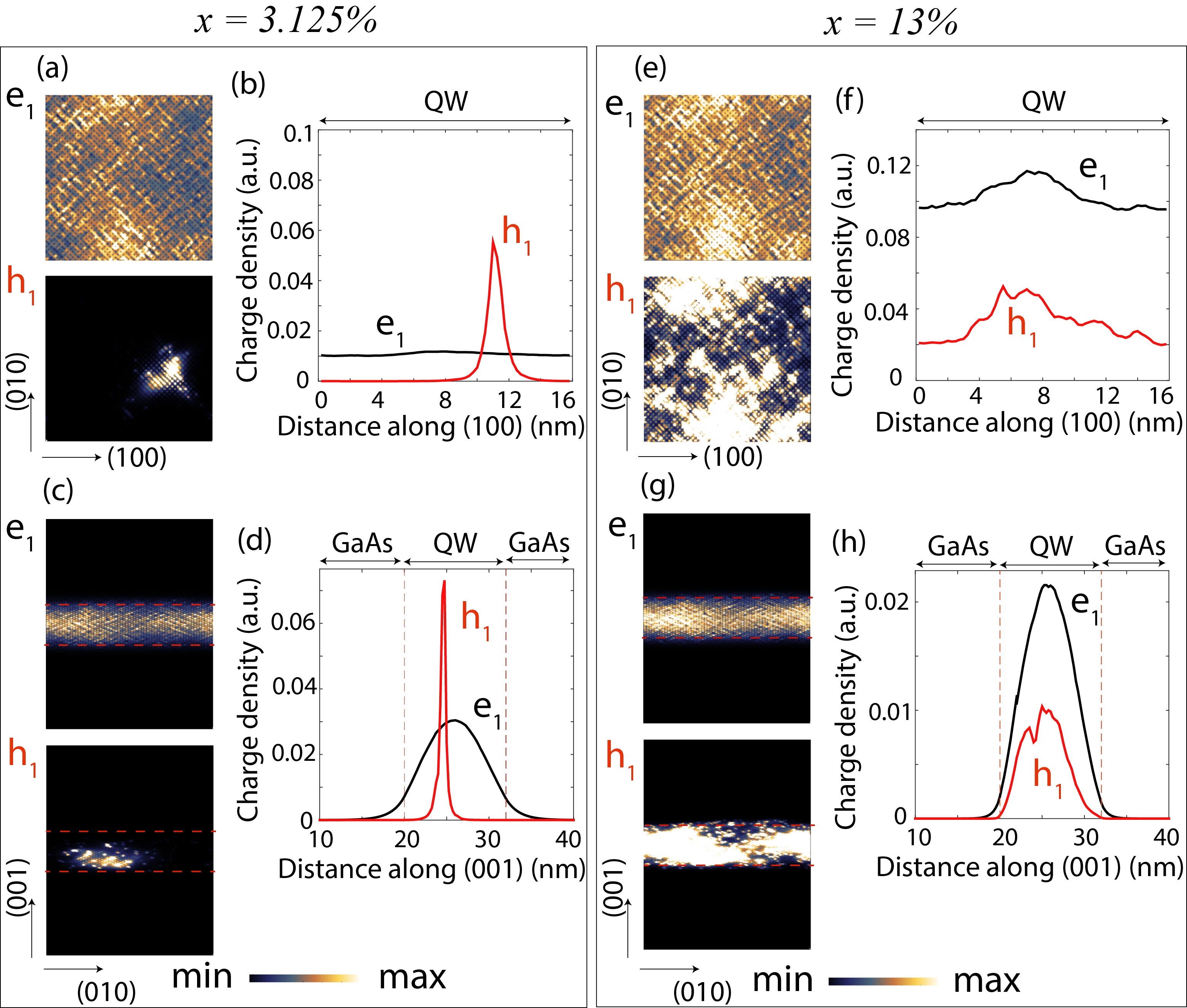}
\caption{(a) A 2D plot of the charge density is shown for the lowest electron (e$_1$) and the highest hole (h$_1$) states along a (001)-plane passing through the center of the QW region with $x$=3.125\%. (b) A line cut plot of the charge density of the e$_1$ and h$_1$ states along the (010)-axis passing through the center of the QW region with $x$=3.125\%. (c) Same as (a) but along a (010)-plane passing through the center of the QW region with $x$=3.125\%. (d) Same as (b) but the line cut is along the (100)-axis passing through the center of the QW region with $x$=3.125\%. (e,f,g,h) The plots are same as (a,b,c,d), but for the QW with $x$=13\%. }
\label{fig:Fig3}
\end{figure*}

\noindent
\textbf{\textit{Strong Confinement of Hole Wave Functions:}} Let's first look at the charge density plots for the QW with $x$=3.125\%. Fig.~\ref{fig:Fig3} (a) shows the 2D color plot of the electron and hole charge densities in a (001)-plane passing through the center of the QW region. The electron charge density (e1) is distributed over the whole QW region although some places of stronger confinement are visible. On the other hand, the hole charge density is very strongly confined and present only inside a small region of the QW. This is also evident from the 1D plots of the charge densities in (b). The charge density plots in other directions shown in (c) and (d) exhibit the similar nature of the confinement. The strong confinement of the hole charge density is attributed to the presence of Bi clusters~\cite{Usman_PRM_2018}. It is well-established that the presence of Bi in GaAs introduces Bi related resonance states within the GaAs valence band, which interact with the valence band edge through a band anti-crossing interaction~\cite{Usman_PRB_2011}. Therefore the presence of Bi clusters strongly impact the confinement of hole charge densities, whereas the electron wave functions remain relatively unperturbed and display envelope wave function character. Recent experimental measurements have provided a clear evidence for the presence of Bi clusters~\cite{Wu_APL_2014, Puustinen_JAP_2013}. We also increased the lateral size of the QW region to 36 nm consisting of 1.33 million atoms and found the presence of similar strong confinement of the hole wave functions at $x$=3.125\% (see Fig. S3 in the supplementary material).  

\begin{figure*}
\includegraphics[scale=0.19]{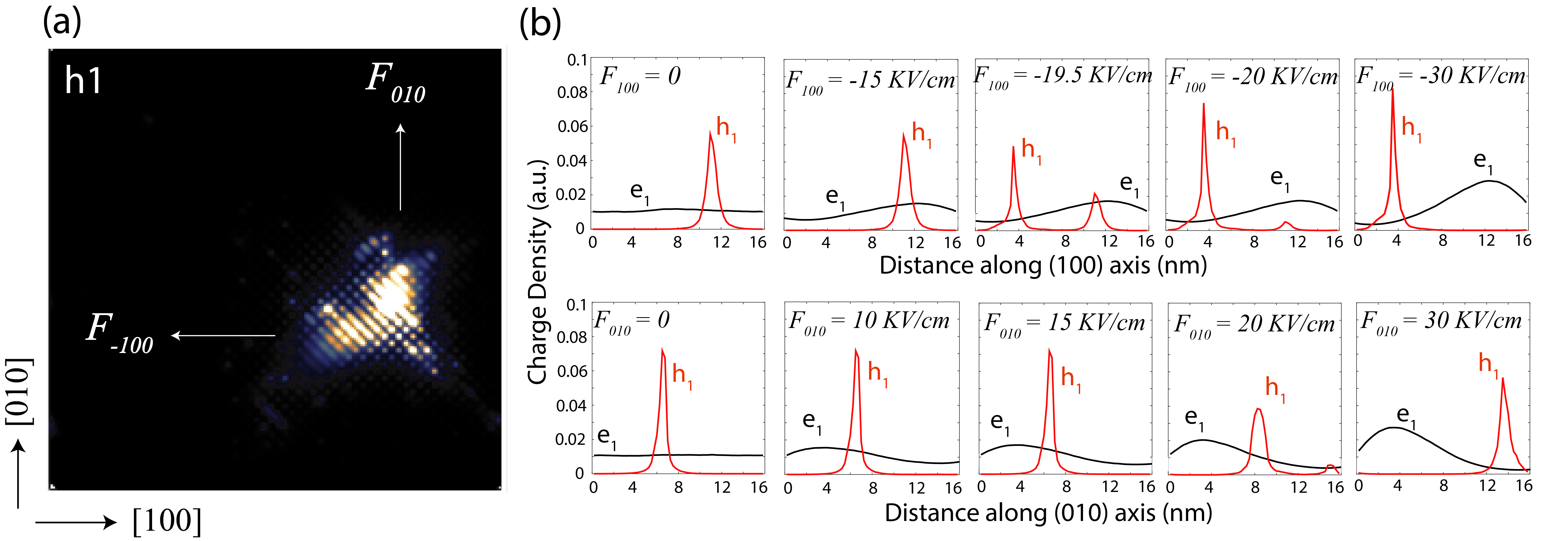}
\caption{(a) A 2D color plot of the highest hole state charge density is shown in a (001)-plane passing through the center of the QW region. (b) Upper row -- The line-cut plot of the lowest electron (e1) and the highest hole (h1) charge densities are shown along the (100)-axis as a function of the applied electric field along the [-100] direction ($\vec{F}_{-100}$). Due to the weak confinement of electron wave function, a charge density shift is computed for $\vec{F}_{-100} \geq$ 0, whereas substantially large electric field magnitude ($\geq$ 20 kV/cm) is required to shift the much more strongly confined hole charge density. Lower row -- Same as the upper row but the line cuts and the applied electric fields are along the [010] axis.}
\label{fig:Fig4}
\end{figure*}

At the larger Bi fraction ($x$=13\%), Fig.~\ref{fig:Fig3} (e) shows that the electron charge density is nearly uniformly distributed over the whole QW region. However somewhat surprisingly, the hole charge density is also significantly spread with a much weaker confinement and roughly looks like the electron charge density. This is also evident from the plots of the charge density in other directions given in (f), (g), and (h) parts of Fig.~\ref{fig:Fig3}. At $x$=13\%, the QW region consists of 12778 Bi atoms (compared to 3072 for 3.125\%), therefore the overall strain in the QW region is much more uniform compared to the $x$=3.125\% case where relatively few Bi atoms implies isolated presence of Bi clusters. The charge density plots for the other two Bi fractions 6.5\% and 10\% are provided in the supplementary material section for reference (Fig. S4). These plots clearly exhibit a trend that the uniformity of the spatial distribution of the hole charge density increases as the Bi fraction of the QW region increases.  

\noindent
\textbf{\textit{Spatial Shift of Hole Charge Density by Electric Field:}} The application of an electric field shifts the electron and hole wave function charge densities in the opposite directions. Ideally if the electron and hole charge densities are symmetric and uniformally distributed in the QW region, the shifts in charge densities will also be symmetric around $|\vec{F}_m|$=0. However typically even at $|\vec{F}_m|$=0, the electron and hole wave functions are separated in a nanostructure (QW or QD) due to the random alloying effect or shape asymmetry of the nanostructure~\cite{Fry_PRL_2000}. This leads to inequivalent shifts in the charge densities where the electrons and holes are differently impacted by the same magnitude and direction of the electric field. Another form of asymmetry in the QCSE has been previously observed, which is linked with the preferred spatial confinement of the electron or hole wave functions that then leads to inequivalent Stark shifts of the corresponding energies depending upon the positive or negative sign of the electric fields ~\cite{Fry_PRL_2000}. We have provided the electric field dependent charge density plots for $x$=6.5\%, 10\%, and 13\% in the supplementary material (Fig. S5), which exhibit increasingly conventional shifts as $x$ increases, previously observed in many other III-V QWs studies. In Fig.~\ref{fig:Fig4}, we only focus on $x$=3.125\% case where the hole charge density is strongly confined in a small region of the QW as shown in (a). The confinement of the hole wave functions towards the bottom right corner of the (001)-plane implies that the application of electric fields $\vec{F}_{-100}$ and $\vec{F}_{010}$ along the (-100) and (010) directions will have to significantly displace the charge density towards the opposite edge as indicated by arrows in Fig.~\ref{fig:Fig4} (a).

Fig.\ref{fig:Fig4} (b) plots the hole charge densities for various magnitudes of the $\vec{F}_{-100}$ and $\vec{F}_{010}$ electric fields. The plots in the upper row shows that the hole wave function (h1) remains confined inside the QW region bounded by 8 nm and 12 nm along the (100) direction until $|\vec{F}_{-100}|$ is increased to 19 kV/cm. The plot of the wave function charge density at $\vec{F}_{100}$ = -19.5 kV/cm shows a hybridized wave function partially confined towards the lower edge of the QW region and the remaining still confined towards the upper edge. The complete shift of the hole wave functions only occurs for $\vec{F}_{100} \leq$ -20 kV/cm. This explains the unusual Stark shifts of the hole energy (h1) plotted earlier in Fig.~\ref{fig:Fig2} (b). To explain it further, we have provided a schematic diagram in the supplementary material section (see Fig. S6). The application of the electric field tilts the valence band edge such that the band edge energy in the 0 $\leq d_{100} \leq$ 8 nm region increases whereas it decreases in the region bounded by 8 $\leq d_{100} \leq$ 16 nm, where $d_{100} $ is the distance along the (100) direction. Since for -20 $\leq \vec{F}_{100} \leq$ 40 kV/cm the hole wave function is confined in the 8 $\leq d_{100} \leq$ 16 nm region, therefore its energy decreases. For -40 $\leq \vec{F}_{100} \leq$ -20 kV/cm, the hole wave function shifts in the 0 $\leq d_{100} \leq$ 8 nm region where the VBE is tilted towards higher energy, so the hole energy immediately increases with a larger slope. 

The same phenomena happens when the electric field direction is along the (010) direction. As shown in the lower row of the Fig.~\ref{fig:Fig4}(b), the hole wave function is now confined in the 0 $\leq d_{010} \leq$ 8 nm region at zero electric field magnitude (Fig. S6). For $\vec{F}_{010} \leq$ 0, the VBE in the 0 $\leq d_{010} \leq$ 8 nm region increases in energy and therefore the hole energy also increases. For $\vec{F}_{010} \geq$ 0, the VBE in the 0 $\leq d_{010} \leq$ 8 nm region decreases and it increases in the 8 $\leq d_{010} \leq$ 16 nm region. In this case, the hole wave function needs to shift from 0 $\leq d_{010} \leq$ 8 nm to 8 $\leq d_{010} \leq$ 16 nm region to increase its energy. However as shown in the Fig.~\ref{fig:Fig4}(b), the hole wave function, due to its very strong confinement, remains confined in the 0 $\leq d_{010} \leq$ 8 nm region until the electric field magnitude is increased above 20 kV/cm. Therefore the hole energy first decreases for 0 $\leq \vec{F}_{010} \leq$ 20 kV/cm and then it starts increasing for  20 $\leq \vec{F}_{010} \leq$ 40 kV/cm.  

Contrarily, the electron wave function exhibits a conventional shift when the electric field is applied: it immediately shifts towards the decreasing side of conduction band edge as soon the electron field magnitude is increased above zero. Therefore the Stark shift of electron energy is quadratic for all directions of the applied electric fields. We also note that the shift in the hole energies is linear due to the stronger confinement, compared to the quadratic shift in e1 energies due to weaker confinement of the corresponding wave functions.

\noindent
\textbf{\textit{Effect of Random Positioning of Bi Atoms:}} In the above sections, we have shown that the unconventional character of the QCSE at $x$=3.125\% arise from a very strong spatial confinement of the hole wave functions, which leads to a highly asymmetric Stark shift in the valence band energies. The strong confinement of the hole wave functions, which is sensitive to the presence of Bi clusters depends on the atomic-level spatial arrangement of the Bi atoms in the QW region. In our calculations, the Bi atoms are placed randomly based on a seed value, therefore different seed values will generate different random spatial distributions of the Bi atoms in the QW region. As a result, the QW region will contain different size and types of Bi clusters. In order to investigate the effect of the spatial distributions of the Bi atoms on the QCSE, we simulated five additional random configurations of the Bi atoms labelled as $RD1$ to $RD5$ in the supplementary figure S6. For each case, we have plotted the spatial distribution of the lowest electron and the highest hole wave functions along the (100) and (010) directions and the QCSE effect for $\vec{F}_{100}$ electric field. Our calculations confirm the presence of the strong confinement of the hole wave functions in all of the five different random distributions of the Bi atoms. The overall variation in the computed band-gap wavelength for the various Bi configurations is $\sim$ 20 nm at zero electric field, which is consistent with the published results at low Bi fraction in the QW region~\cite{Usman_APL_2014}. On the other hand, as a function of the applied electric field, we find that the change in the band-gap is $\sim$ 30 nm. Interestingly, this change is comparable to the variation related to the alloy randomness at 3.125\% Bi fraction.

The plots of the QCSE in supplementary Fig. S6 indicate that the quadratic to linear change in the wavelength dependence on the applied electric field is valid for all the investigated cases when the hole wave functions are confined close to the edges of the QW regions. For the $RD2$ configuration, the hole wave function (h1) exhibits a strong confinement close to the center of the QW region. In this case, our results show that the QCSE is linear for the whole range of the applied electric field, implying that the nature of QCSE (linear or quadratic) not only depends on the strong confinement of the hole wave functions but also on the relative spatial location inside the QW region. The linear QCSE for this particular case can be understood based on the presence of the hole wave function close to the centre of QW region, which implies that the application of a small non-zero electric field is sufficient to shift the hole wave function in either direction. This is the reason for the minimum value of the GSTW being at $\vec{F}_{100}$=0. Finally, we have provided the fitted values of $\alpha$ and $\beta$ in Fig. S6 for all of considered random configurations. Overall the value of $\alpha$ varied from 500$\times$10$^{-6}$ eV.cm/kV to 800$\times$10$^{-6}$ eV.cm/kV, and the value of $\beta$ varies from -5$\times$10$^{-6}$ (eV.cm/kV)$^2$ to -7$\times$10$^{-6}$ (eV.cm/kV)$^2$ where the QCSE is quadratic. We should point out here that these variations are computed at Bi=3.125\%, for which the impact of alloy configurations is strongest. As the Bi fraction in the QW region increases, the alloy randomness effect becomes weak and therefore we expect significantly smaller variation in $\alpha$ and $\beta$ at Bi=13\%.

\begin{figure*}
\includegraphics[scale=0.3]{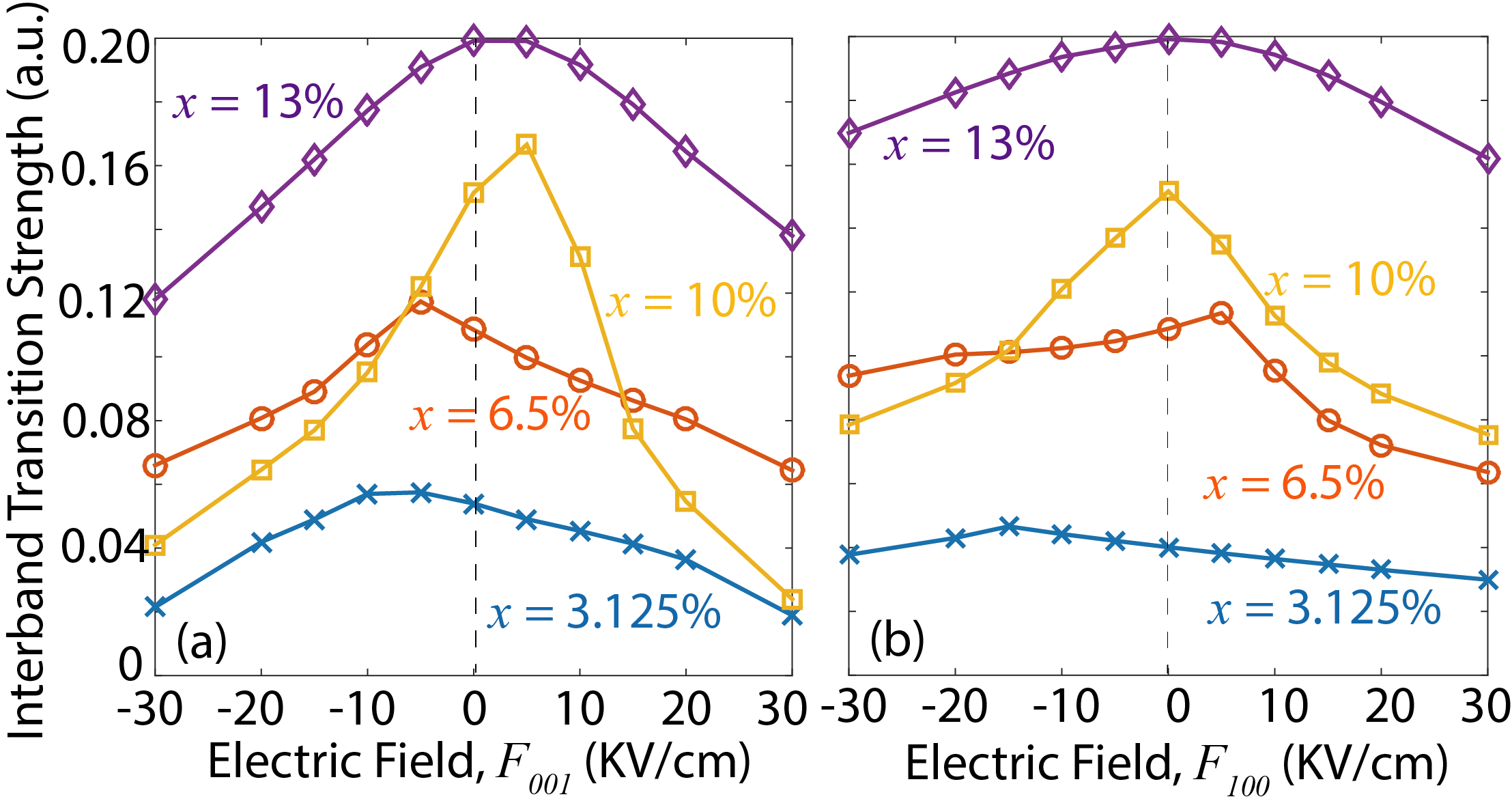}
\caption{The plots of inter-band optical transition strengths are shown for $x$= 3.125\%, 6.5\%, 10\%, and 13\% as a function of the applied electric fields along the (a) [001] axis and (b) [100] axis.}
\label{fig:Fig5}
\end{figure*}

\noindent
\textbf{\textit{Inter-band Optical Transition Strengths:}} Fig.~\ref{fig:Fig5} plots the optical transition strengths (OTS) as a function of the applied electric fields for various Bi fractions in the QW region. The momentum matrix element between the ground electron and hole states is computed as follows~\cite{Usman_PRB2_2011}:
\noindent
\begin{eqnarray}
M_{\overrightarrow{n}}^{\alpha \beta} &=& \sum_{i,j} \sum_{\mu,\nu} (C^e_{i,\mu,\alpha})^* (C^h_{j,\nu,\beta}) {\langle i\mu\alpha \vert \textrm{\textbf{H}} \vert j\nu\beta \rangle} {(\overrightarrow{n}_{i}-\overrightarrow{n}_{j})} \label{eq:momentum_x}
\end{eqnarray}
\noindent
where $\alpha$ and $\beta$ indicates the spin up and down states, \textbf{H} is the tight-binding Hamiltonian in \spss basis, and $\overrightarrow{n} = \overrightarrow{n}_{i} - \overrightarrow{n}_{j}$ is the spatial distance between atoms $i$  and $j$. For TE mode OTS calculation, $\overrightarrow{n}$ is equal to $\overrightarrow{x}_{i} - \overrightarrow{x}_{j}$ and it is equal to $\overrightarrow{z}_{i} - \overrightarrow{z}_{j}$ for the TM mode OTS calculation. After calculting the momentun matrix elements, we compute the TE and TM mode optical transition strengths (TE$_{100}$ and TM$_{001}$) from Fermi's Golden rule (by summation of the absolute values of the momentum matrix elements including spin degenerate states)~\cite{Usman_PRB2_2011}:
\noindent
\begin{eqnarray}
\textrm{TE$_{100}$} &=&  \sum_{\alpha, \beta} \vert M_{\overrightarrow{x}}^{\alpha \beta} \vert ^2 \label{eq:TE_X} \\
\textrm{TM$_{001}$} &=&  \sum_{\alpha, \beta} \vert M_{\overrightarrow{z}}^{\alpha \beta} \vert ^2  \label{eq:TM_Z}
\end{eqnarray}   

\noindent
The OTS between the ground-state electron and hole states is strongly dependent on the overlap between the corresponding wave functions. The application of an electric field shifts the electron and hole wave functions in the opposite directions, leading to a decrease in the spatial overlap between them. This results in a decrease in the OTS magnitude as a function of the electric field. This is true for all the Bi fractions and for both electric field orientations plotted in Fig.~\ref{fig:Fig5}. The plots also exhibit two general trends for the OTS. The first is as a function of the Bi fraction $x$: the magnitude of the OTS increases when $x$ increases which is true for both $\vec{F}_{100}$ and $\vec{F}_{001}$ fields. This is because the hole wave function is highly confined at low Bi fractions and it results in lesser overlap with the relatively uniformly confined electron wave function. The second trend is related to increase in the electric field magnitude, which leads to an overall larger decrease in the OTS magnitude at higher Bi fractions when compared to the low Bi fractions. This is related to the electric field induced shift in the hole wave functions, which is stronger due to the weaker confinement level of electrons and holes. As the shift in electron wave functions is relatively similar for all $x$, a larger decrease in the electron-hole wave function spatial overlap leads to a correspondingly larger decrease in the OTS magnitude for $x$=10\% and 13\%.    
    
\section{Conclusions}
\noindent
Based on large-scale atomistic tight-binding simulations, we have investigated the quantum confined Stark effect (QCSE) for novel \GaBiAsGaAs quantum wells. To provide a comprehensive understanding of the QCSE, we have varied the magnitude and orientation of the electric field for four different Bi compositions $x$= 3.125\%, 6.5\%, 10\%, 13\%. Our results have revealed that at low Bi fractions ($x$=3.125\%), the character of the QCSE is dominated by the strong confinement of the hole wave functions due to the presence of Bi clusters. This leads to an unconventional nature of the QCSE. At large Bi compositions, the hole wave function confinement is weakly dictated by the presence of the Bi clusters, and therefore the QCSE exhibits a conventional quadratic dependence on the electric field irrespective of its orientations. At technologically relevant 1300 nm wavelength reported for the 13\% Bi fraction, we have computed an overall shift of 22 nm in the GSTW based on a 40 kV/cm variation in the magnitude of the electric field, irrespective of its orientation. Our results document the first study of the QCSE for technologically relevant \GaBiAsGaAs QWs and provide useful guidance to design optoelectronic and spintronic devices. 

\section{Acknowledgements}
\noindent
Computational resources are acknowledged from National Computational Infrastructure (NCI) under the National Computational Merit based Allocation Scheme (NCMAS).   

\bibliographystyle{elsarticle-num}

\clearpage
\newpage

\Large \textbf{\textit{\underline{Supplementary Figures}}}
\\ \\

\renewcommand{\thefigure}{S\arabic{figure}}
\setcounter{figure}{0}

\renewcommand{\theequation}{S\arabic{equation}}
\setcounter{equation}{0}

\begin{figure*}
\centering
\includegraphics[scale=0.4]{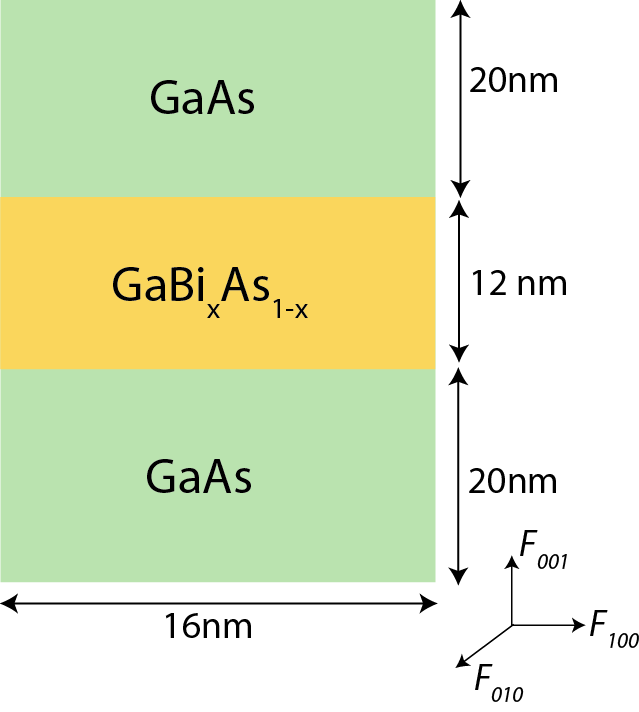}
\caption{Schematic diagram of a \GaBiAsGaAs quantum well structure is illustrated. The device consists of a single \GaBiAsGaAs QW with 12 nm width along the confinement direction (001). The thickness of both GaAs substrate and capping layer is 20 nm, and the lateral dimensions are 16 nm in each of the (100) and (010) directions. The boundary conditions are periodic in all three spatial dimensions.}
\label{fig:FigS1}
\end{figure*}

\begin{figure*}
\includegraphics[scale=0.25]{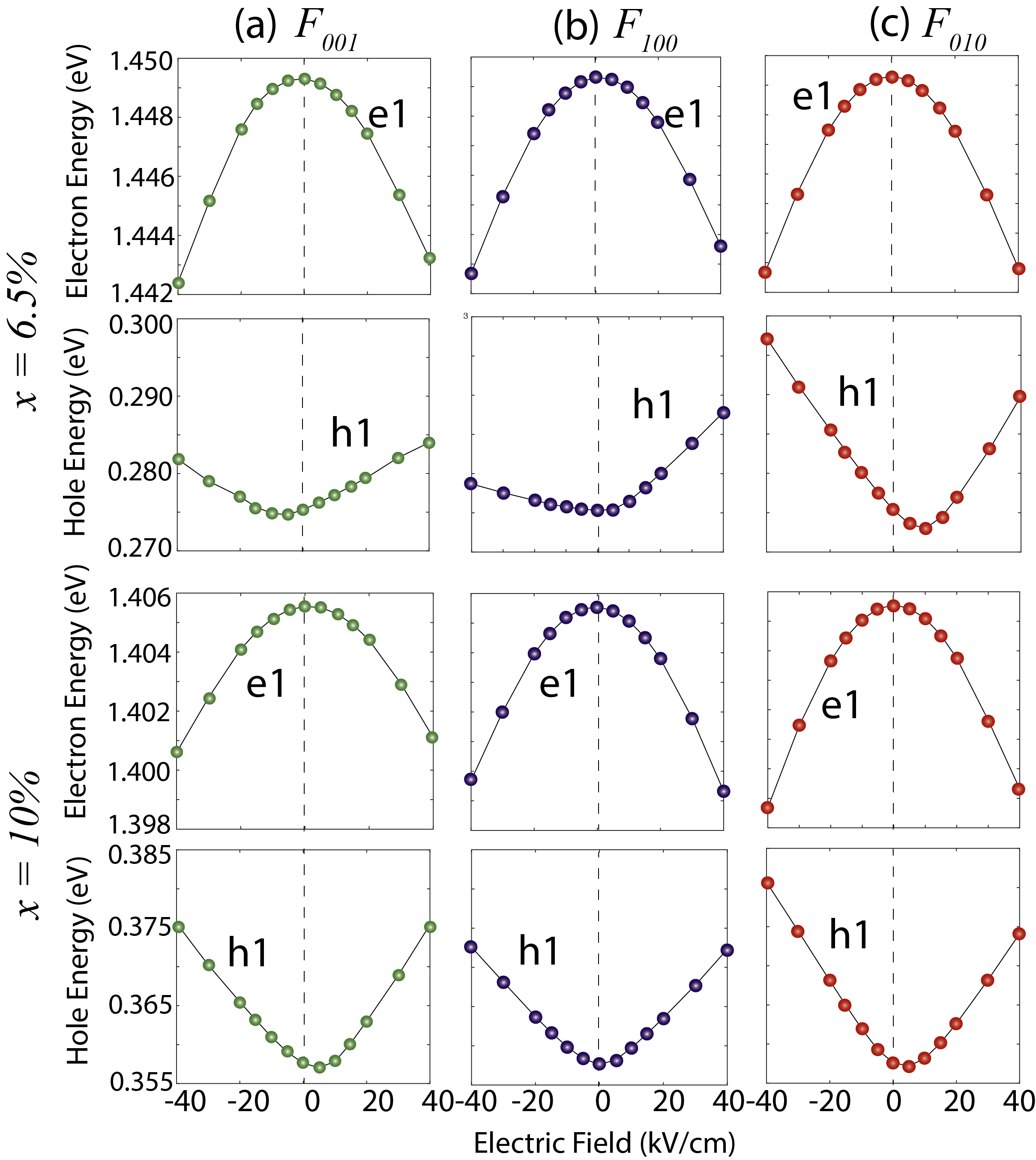}
\caption{The plots of the lowest electron (e1) and the highest hole (h1) energies are shown for Bi factions 6.5\% and 10\% in the QW region as a function of the applied electric fields along the (a) (001), (b) (100), and (c) (010) axes.}
\label{fig:FigS2}
\end{figure*}

\begin{figure*}
\includegraphics[scale=0.4]{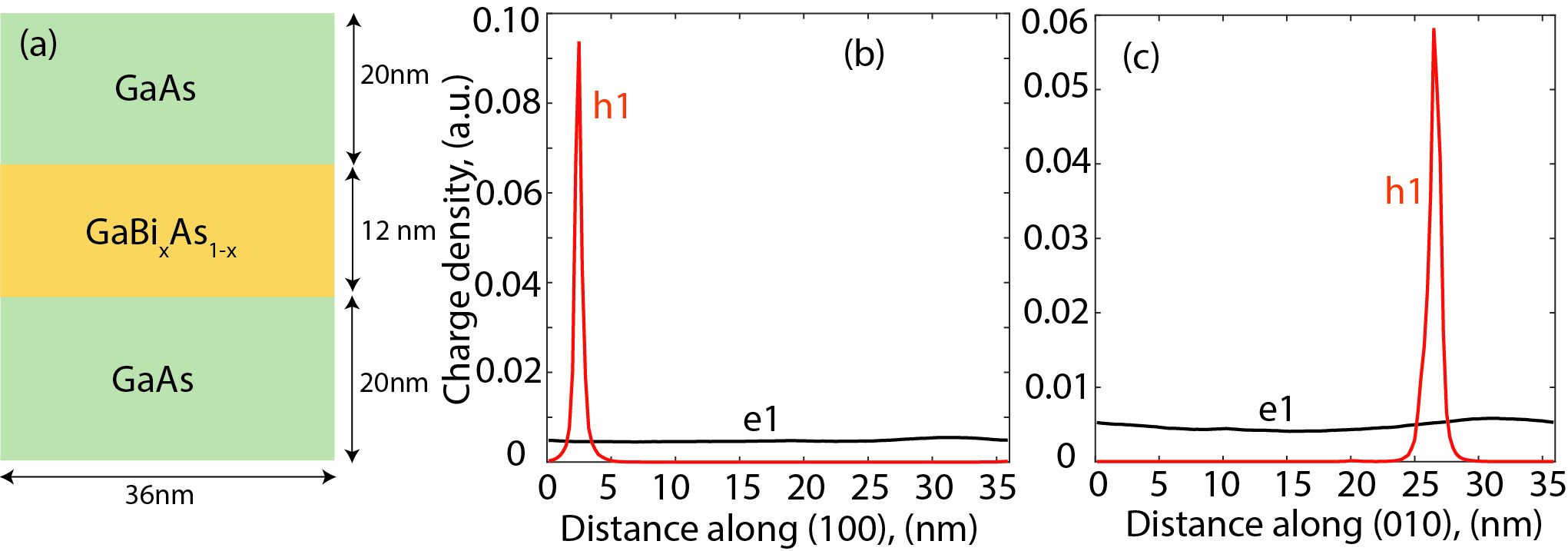}
\caption{(a) Schematic diagram of a \GaBiAsGaAs quantum well structure is illustrated. The device consists of a single \GaBiAsGaAs QW with 12 nm width along the confinement direction (001). The thickness of both GaAs substrate and capping layer is 20 nm, and the lateral dimensions are 36 nm in each of the (100) and (010) directions. The boundary conditions are periodic in all three spatial dimensions. (b, c) Line plots of the lowest electron (e1) and the highest hole (h1) charge densities are shown along the (100) and (010) axes, passing through the center of the QW region. }
\label{fig:FigS3}
\end{figure*}

\begin{figure*}
\includegraphics[scale=0.18]{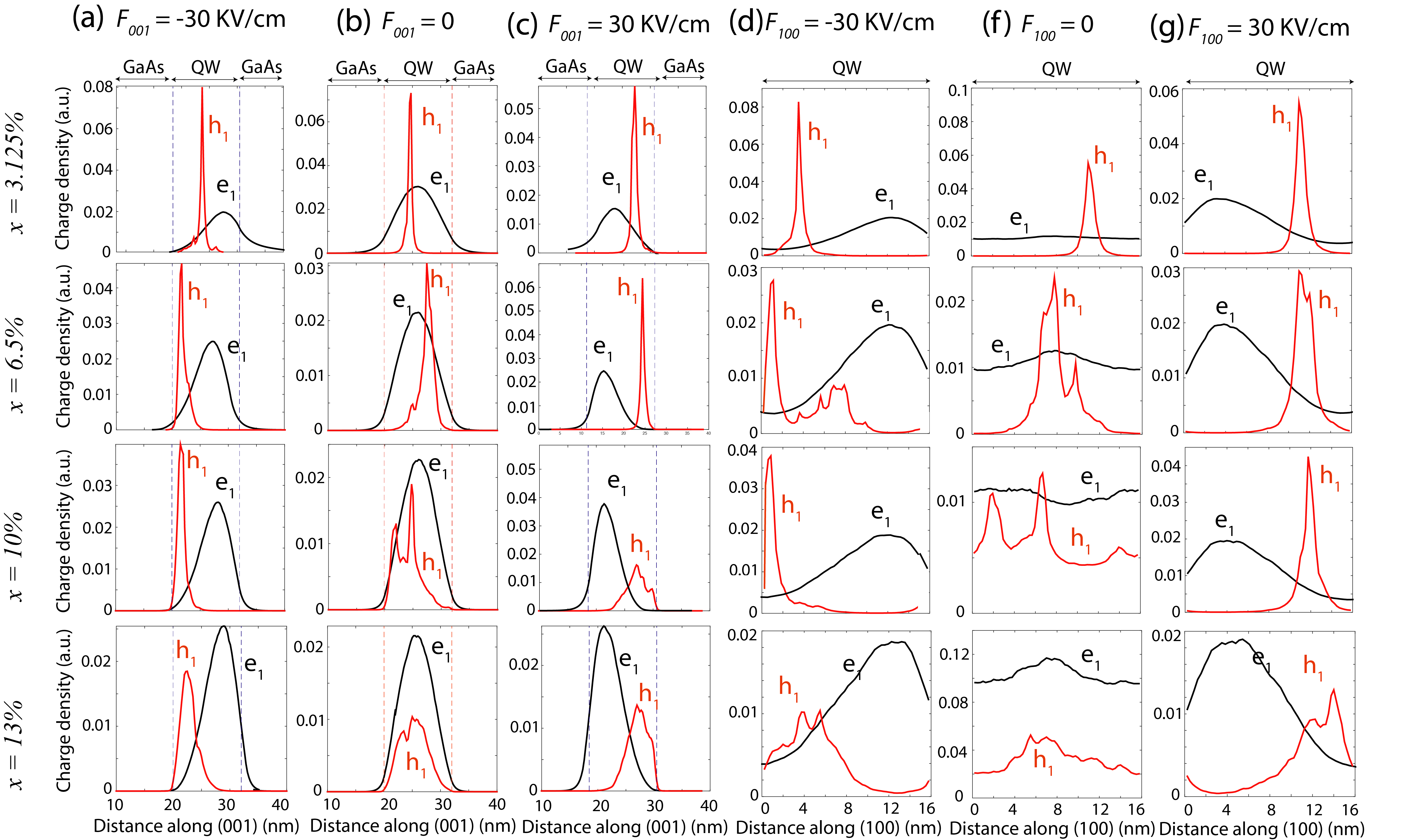}
\caption{The plots of the lowest electron (e1) and the highest hole (h1) charge densities are shown as a function of the distance along the (001) and (100) axis. In each plot, the line cut passes through the center of the QW region. The columns (a) to (g) represent the applied electric field orientation and magnitude. The rows are different Bi compositions of the QW region.}
\label{fig:FigS4}
\end{figure*}

\begin{figure*}
\includegraphics[scale=0.4]{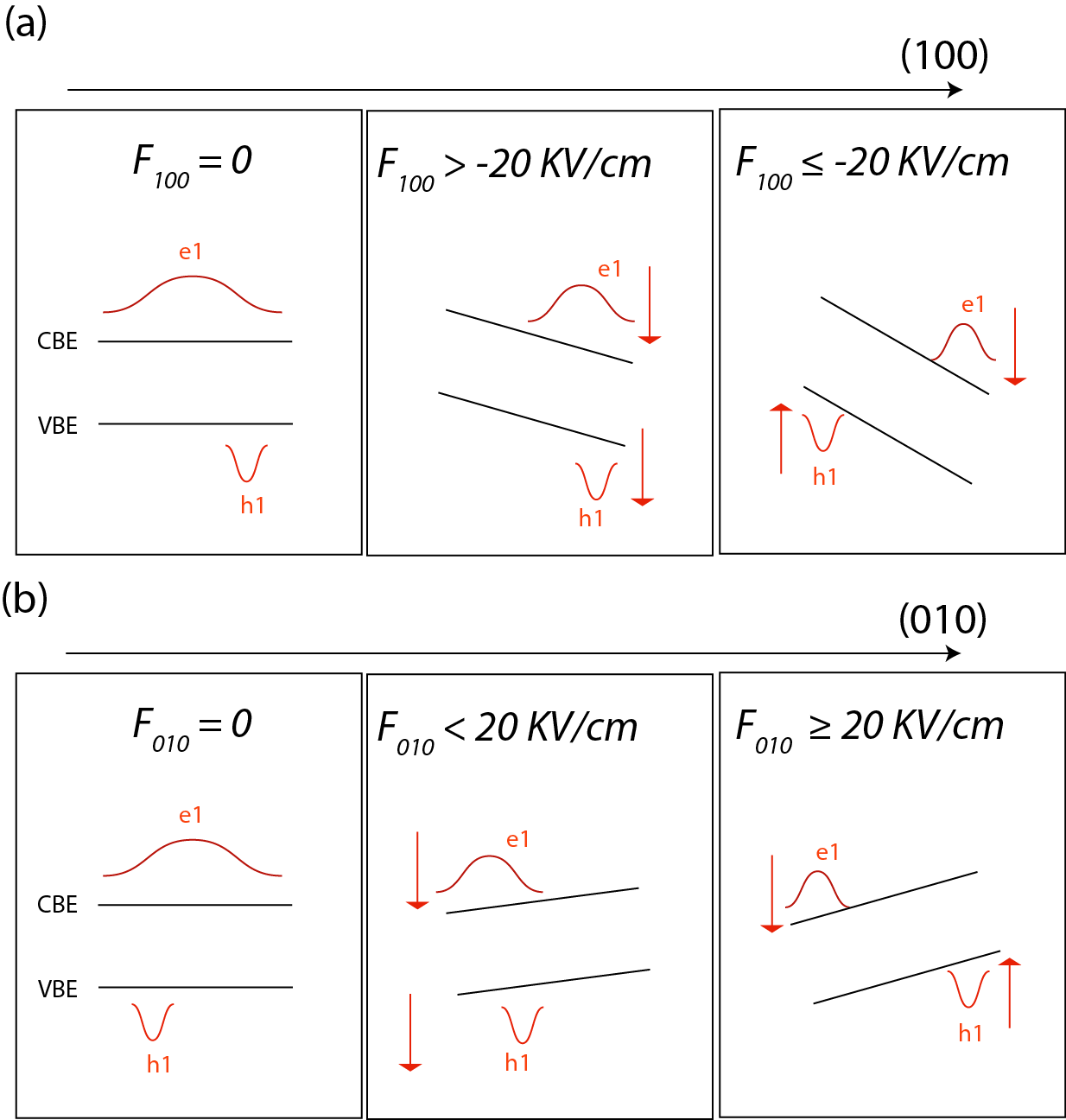}
\caption{(a) This is the schematic diagram which shows the impact of the applied electric field on the lowest conduction band edge (CBE) and the highest valence band edge (VBE) along the in-plane (100) direction. The positions of the confined electron and hole wave functions are also indicated. For $F_{100} \geq$ -20 kV/cm, the hole wave function is confined towards the right edge of the QW region and it experiences a decease in the energy. For $F_{100} \leq$ -20 kV/cm, the hole wave function is confined towards the left edge of the QW region and therefore its energy increases. (b) Same as (a) but for the (010) axis and $F_{010}$ field orientation. }
\label{fig:FigS5}
\end{figure*}

\begin{figure*}
\includegraphics[scale=0.3]{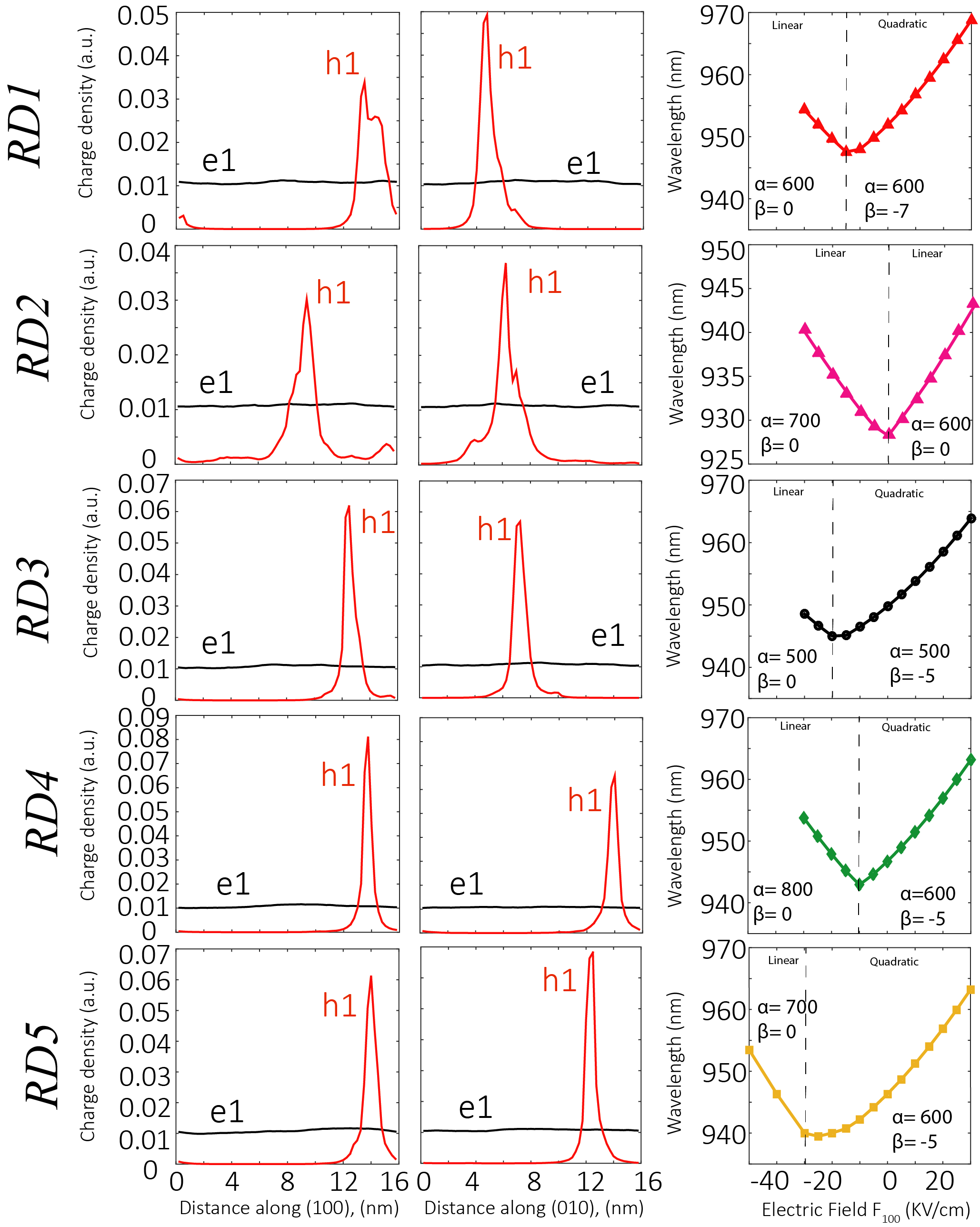}
\caption{The data in this figure corresponds to 3.125\% Bi fraction in the QW region. (a) The line cut plots of the lowest electron (e1) and the highest hole (h1) charge densities are shown along the (100) axis passing through the center of the QW region. (b) The same as (a) but the line cut passes through the center of the QW region along the (010) axis. (c) The plots of the ground state transition wavelength (GSTW) are shown as a function of the applied electric field along the (100) direction. Note that the five rows indicate five different random configurations of Bi atoms inside the QW region labelled as RD1, RD2, RD3, RD4, and RD5. The plots of the wavelength as a function of electric field also include the fitted values of $\alpha$ in the units of eV.cm/KV and $\beta$ in the units of (eV.cm/KV)$^2$, in accordance with the equation 1 of the main text.}
\label{fig:FigS6}
\end{figure*}

\end{document}